\documentclass[pre,twocolumn,showpacs,floatfix,superscriptaddress,aps]{revtex4}
\usepackage{graphicx}
\usepackage{amsmath,amssymb}

\begin{document}

\title{Analytical calculation of the transition to complete phase synchronization in coupled oscillators}

\author{Paulsamy Muruganandam}

\affiliation{School of Physics, Bharathidasan University, Palkalaiperur, Tiruchirappalli -- 620024, India}

\affiliation{Instituto de F\'{\i}sica Te\'orica, Universidade Estadual Paulista, R.  Pamplona 145, 01405-900 S\~ao Paulo, Brazil}

\author{Fernando F. Ferreira}

\affiliation{Grupo Interdisciplinar de F\'{\i}sica da Informa\c{c}\~ao e Economia (GRIFE), Escola de Arte, Ci\^encias e Humanidades, Universidade de S\~ao Paulo,  Av. Arlindo Bettio 1000, 03828-000 S\~ao Paulo, Brazil}

\author{Hassan F. El-Nashar}

\affiliation{Department of Physics, Faculty of Science, Ain Shams University, 11566 Cairo, Egypt}

\affiliation{Department of Physics, Faculty of Education, P.O.~Box 21034, 11942 Alkharj, K.S.A}

\author{Hilda A. Cerdeira}

\affiliation{Instituto de F\'{\i}sica Te\'orica, Universidade Estadual Paulista, R.  Pamplona 145, 01405-900 S\~ao Paulo, Brazil}

\affiliation{Instituto de F\'{\i}sica, Universidade de S\~ao Paulo, R. do Mat\~ao, Travessa R. 187, 05508-090 S\~ao Paulo, Brazil}


\begin{abstract}

Here we present a system of coupled phase oscillators with nearest neighbors coupling, which we study for different boundary conditions. We concentrate at the transition to total synchronization. We are able to develop exact solutions for the value of the coupling parameter when the system becomes completely synchronized, for the case of periodic boundary conditions as well as for an open chain with fixed ends. We compare the results with those calculated numerically.

\end{abstract}


\maketitle

\section{Introduction}

The fact that systems of coupled oscillators can describe problems in 
physics, chemistry, biology, neuroscience and many other disciplines, is already widely accepted by the scientific community. They have been used to model diverse phenomena such as: Josephson junction arrays, multimode lasers, vortex dynamics in fluids, biological information processes, neurodynamics \cite{pik, dom, per, kur, win, st1, para, wie}. The coupled oscillators have been observed to synchronize themselves (say to a common frequency value) in a variety of ways, such as total, partial, generalized, lag, etc., when the coupling strength varies \cite{gra, kan, had,  lor, ot, h1, h2, h3}. In spite of the diversity of dynamics, all these systems synchronize themselves to a common frequency, in a tree-like clustering behavior, when the coupling strength between these oscillators increases  \cite{hak, dp,tass, boc1, pik2, wu, gol}. Among the coupled oscillators systems, is the system of coupled limit cycle
 s, which has attracted interests. The synchronization phenomena of such system, in spite of the diversity of dynamics  can be described using simple models of coupled phase equations.

Although there has been an extensive exploration of the dynamical behavior of the coupled limit cycles that show the synchronization phenomena, many interesting features remain unknown and little has been done analytically. Special attention has been given to the complex synchronization tree of a system of non-chaotic oscillators with nearest neighbors sinusoidal interactions, which becomes chaotic when the interaction is turned on \cite{hu2, xx, me, me1, me2}. This system, a diffusive version of the Kuramoto model \cite{kur}, in spite of its simplicity, possesses all the features of phase synchronization of a system of chaotic oscillators.
\begin{align}
\dot{\theta}_i=\omega_i+\frac{K}{3}[\sin(\theta_{i+1}-\theta_i) + \sin(\theta_{i-1} - \theta_i)]. \label{kura:theta}
\end{align}
Here $\omega_i$ are the natural frequencies, selected randomly from a normal Gaussian distribution, $K$ is the coupling strength, $\theta_i$ is the instantaneous phase, $\dot{\theta_i}$ is the instantaneous frequency and $i=1,2,\ldots,N$. Such oscillators with nearest neighbors interaction have been seen in the literature to describe Josephson junctions, laser arrays and phase-locked loops \cite{dp,xx}. These nonidentical oscillators cluster in time averaged frequency, until they completely synchronize to a common value of the average frequency \cite{hu2, me, me1, me2, liu}, $\langle \dot{\theta_i} \rangle = \langle \dot{\theta_j} \rangle$, $i \neq j$, where
\begin{align}
\langle \dot{\theta_i} \rangle =\lim_{T \to \infty} \frac{1}{T} \int_0^T \dot{\theta_i}(t)dt.
\end{align}
Using periodic boundary conditions $\theta_{i+N}=\theta_i$, and scaling the frequencies such that:
\begin{align}
\frac{1}{N}\sum_{i=1}^N\omega_i=0, \label{omega_scale}
\end{align}
the above system (1) of $N$ oscillators has a critical coupling strength $K=K_c$, where at $K \ge K_c$ a complete frequency synchronization can be observed and each $\theta_i$ is locked to a fixed value. There is no phase locking for $K<K_c$ although the system has clusters of oscillators of the same \emph{average} frequencies. Reducing further the coupling strength $K$, the number of units which cluster to the same average frequency decreases, increasing the number of clusters, i.e. the number of branches in the synchronization tree, until finally all oscillators are independent and acquire their natural frequencies.

The scaling of frequencies (\ref{omega_scale}) in the case of system (\ref{kura:theta}) limits the synchronization to zero frequency. When this scaling is not used \cite{me, me1, liu}, the general features of the system and the value of $K_c$ will not change, the synchronization occurs via the same transition tree but shifted to a common frequency value
\begin{align}
\omega_0 = \frac{1}{N}\sum_{i=1}^{N}\omega_i,
\end{align}
as long as we maintain the periodic boundary conditions. When we lift them, the system becomes dependent on the value of $\omega_0$ \cite{me2}, therefore, we shall keep $\omega_0$ different from zero throughout this work.

\begin{figure}[!ht]
\centering\includegraphics[width=\linewidth,clip]{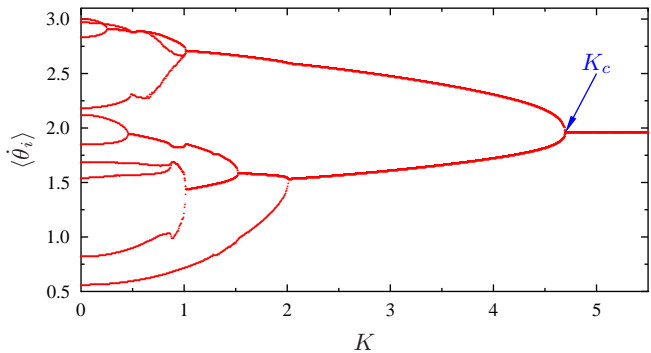}
\caption{(Color online) Synchronization tree for a system of $N=10$ oscillators, with periodic boundary conditions ($K_c \sim 4.70$).}
\label{kura_n10}
\end{figure}

\begin{figure}[!ht]
\centering\includegraphics[width=\linewidth,clip]{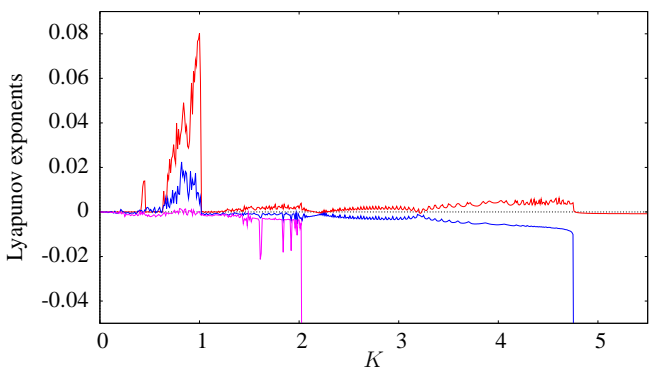}
\caption{(Color online) The first three largest Lyapunov exponents for  a system with $N=10$ oscillators}
\label{lyap_periodic}
\end{figure}

In Fig.~\ref{kura_n10} we show the synchronization tree for a periodic system with $N=10$ oscillators. Also in Fig.~\ref{lyap_periodic} we plot the first three largest Lyapunov exponents as a function of the coupling strength, $K$.  At and above the critical coupling strength, $K_c$, the largest Lyapunov exponent becomes zero with the rest 
of them being negative, an indication of synchronization. The initial frequencies for Figs.~\ref{kura_n10} and \ref{lyap_periodic} are chosen as $\omega_i = \{0.5331$, $2.8577$, $2.6978$, $2.0773$, $2.8257$, $1.6062$, $0.7788$, $1.4680$, $1.7622$, $2.0190\}$.

The analysis of synchronization and the loss of stability becomes particularly difficult since, as we mentioned above, it is on long time average that these systems synchronize. Therefore, for a system (1), one can performs analytic calculations only where the variables (phase and frequency) become time independent. In order to do that, it is necessary to make sure that we are in a zone of stationary states so that averages are equal to instantaneous values of the phases and frequencies. Henceforward, it becomes clear that we need to arrive at $K_c$ from above ($K \ge K_c$).

\section{Method of Lagrange multipliers}
\label{sec2}

\subsection{Periodic Boundary Conditions}

We shall start with the case of periodic boundary conditions. At complete synchronization all average frequencies are equal and all phase differences are constant, while its sum equals zero. Since $\omega_0$ is different from zero it is necessary to use the phase differences defined as $\phi_i = \theta_{i} - \theta_{i-1}$. Then, the equation (\ref{kura:theta}) can be rewritten as
\begin{align}
\dot \phi_i = \omega_{i} - \omega_{i-1} + \frac{K}{3} \left( \sin \phi_{i-1}  - 2 \sin \phi_{i} + \sin \phi_{i+1} \right) \label{kura:phase_diff}
\end{align}
In order to have complete frequency synchronization, the above systems of equations should have stable steady states. It can be seen from Equation (\ref{kura:phase_diff}) that the stability of the frequency steady states is independent of the initial frequencies as well as the coupling constant $K$ provided $K>0$. Thus, if we could establish a criteria for the existence of a stable steady states for a minimum value of $K > 0$, then simply the minimum value of $K$ is equal to the critical coupling strength, $K_c$, since the system remains synchronized once full synchronization has been achieved (for $K \ge K_c$). To find this minimum we use the method of Lagrange multipliers.

We should note that the frequencies of the system synchronize on time 
average, and after complete frequency synchronization all $\dot \phi_i = 0$, and the $\phi_i$'s are all constants. Therefore as long as the calculation is done above $K_c$, we can exchange the variables by their time independent values. It is straightforward to see from the steady states of (\ref{kura:phase_diff}), where $\dot \phi_i=0$, that
\begin{align}
K = \frac{a_2}{\sin \phi_2^*-\sin \phi_1^*}, \label{eq:kval}
\end{align}
and
\begin{align}
\sin \phi_1^* - \sin \phi_2^*  + \frac{a_i}{a_2}(\sin \phi_i^*-\sin \phi_1^*) = 0,\;\; i = 3,4,\ldots,N, \label{eq:cond1}
\end{align}
where
\begin{align}
a_2 = 3 (\omega_0 -\omega_1), \notag
\end{align}
and
\begin{align}
a_i = a_{i-1} +  3 (\omega_0 -\omega_{i-1}), \;\;\; i = 3,4,\ldots, N.
\end{align}
In addition, from the definition of $\phi_i$, it follows that
\begin{align}
\sum_{i=1}^{N} \phi_i^* = 0, \label{eq:cond2}
\end{align}
We optimize the function (\ref{eq:kval}) subject to the conditions (\ref{eq:cond1}) and (\ref{eq:cond2}). Then the minimum value of $K > 0$ is the critical value $K_c$, provided that the set $\{\phi_i^*, \;\;i=1,2,\ldots, N\}$ form a stable steady state of (\ref{kura:phase_diff}).

For the above purpose, let us define a function $E(\phi_1^*, \phi_2^*, \ldots^*, \phi_N^*, \lambda_0, \lambda_3, \lambda_4, \ldots, \lambda_N)$ as
\begin{widetext}
\begin{align}
E = \frac{a_2}{\sin \phi_2^*-\sin \phi_1^*} + \sum_{i=3}^{N} \lambda_i \left[ \sin \phi_1^* - \sin \phi_2^*  + \frac{a_i}{a_2}(\sin \phi_i^*-\sin \phi_1^*) \right] + \lambda_0 \sum_{i=1}^{N} \phi_i^*,
\end{align}
\end{widetext}
where $\lambda_i$, $i=3,4\ldots,N$ are parameters. To have an optimum $K$ one should solve
\begin{align}
& \frac{\partial E}{\partial \phi_i^*} = 0, \;\;\; i = 1,2,3, \ldots, N \notag \\
& \frac{\partial E}{\partial \lambda_i} = 0, \;\;\; i = 3, 4, \ldots, N
\end{align}
The above conditions yield the following algebraic equation in addition to Eqs.~(\ref{eq:cond1}) and (\ref{eq:cond2})
\begin{align}
\sum_{i=1}^{N} \frac{\cos\phi_1^*}{\cos\phi_i^*} = 0, \label{eq:cond3}
\end{align}
We can obtain the steady states $\phi_i^*$, $i=1,2,\ldots, N$ values by solving Eqs.~(\ref{eq:cond1}), (\ref{eq:cond2}) and (\ref{eq:cond3}).
Hence we solve the above set of equations using Newton-Raphson method. Assuming a random set of initial values for $\phi_i^*$'s, we look for converged and stable values for $\phi_i^*$'s, then the values of $K>0$ are estimated using Eq.~(\ref{eq:kval}). The application of these equations bring the results shown in Table~\ref{table1}.
\begin{table}[!ht]
\caption{Critical $K$ values ($K_c$) for the case periodic boundary conditions}
\label{table1}
\begin{tabular}{crr}
\hline
 \multicolumn{1}{c}{~~N~~}
    & \multicolumn{2}{c }{$K_c$}  \\
 \multicolumn{1}{c}{}
    & \multicolumn{1}{c}{Numerical\footnote{from direct numerical simulation of
    ({\ref{kura:theta})}}}
    &  \multicolumn{1}{c}{Analytical\footnote{by solving
    Eqs.~(\ref{eq:cond1}), (\ref{eq:cond2}) and (\ref{eq:cond3})}}  \\
\hline
  4 &  2.540775~~ & 2.540771~~  \\
  5 &  2.535587~~ & 2.535583~~  \\
  6 &  3.293280~~ & 3.293245~~  \\
  7 &  4.561624~~ & 4.561588~~  \\
 10 &  4.696059~~ & 4.696048~~  \\
 15 &  2.851727~~ & 2.851651~~  \\
 25 &  3.642530~~ & 3.642538~~  \\
 50 &  9.458391~~ & 9.458378~~  \\
100 &  12.723401~~ & 12.723417~~  \\
\hline
\end{tabular}
\end{table}

From Table~\ref{table1}, we notice that the analytically and numerically calculated values are in a good agreement.

Recently, Daniels et al. \cite{xx} presented a calculation of equation (5) to estimate the value of the critical coupling $K_c$ based on the results by Strogatz and Mirollo \cite{st1}. Our method, which makes use of the fact that above total synchronization ($K \ge K_c$) all variables: phases and frequencies, are time independent and the problem of handling temporal averages to estimate the value of the critical coupling $K_c$, disappears. In addition, we are able to obtain a condition for complete synchronization to occur [equations (\ref{eq:cond1}), (\ref{eq:cond2}) and (\ref{eq:cond3})].

\subsection{Fixed Ends}

\begin{figure}[!ht]
\centering\includegraphics[width=\linewidth,clip]{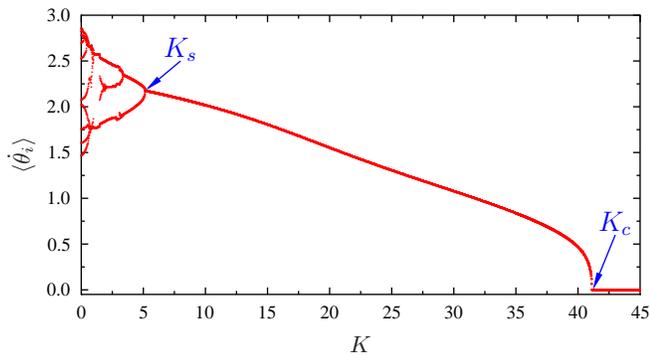}
\caption{(Color online) Synchronization tree for a system of $N=10$ oscillators, with fixed ends ($K_s \sim 5.18$ and $K_c \sim 41.13$).}
\label{kura_n10_fix}
\end{figure}

We move now to the system where the phases at the boundaries are fixed. In this case, with the boundary conditions $\theta_N=c_N$ and $\theta_0=c_0$, we can write the following equations:
\begin{subequations}
\label{eqn:fixed}
\begin{align}
\dot \theta_{N+1}&=0\\
\dot \theta_i &= \omega_i + \frac{K}{3} \left[\sin\left(
\theta_{i-1}-\theta_{i} \right) + \sin\left(
\theta_{i+1}-\theta_{i} \right)\right] \\
\dot \theta_{0}& = 0
\end{align}
\end{subequations}
where $i = 1,2,\ldots, N$.
The equation (\ref{eqn:fixed}) can be written in terms of the phase differences $\phi_i = \theta_i - \theta_{i-1}$ as
\begin{align}
& \dot \phi_1 = \omega_1 + \frac{K}{3} \left( \sin \phi_{2}  -  \sin \phi_{1} \right), \notag \\
& \dot \phi_i = \omega_{i}-\omega_{i-1}+\frac{K}{3}\left(\sin\phi_{i-1}-2\sin\phi_{i}+ \sin \phi_{i+1} \right),\notag \\
& \dot \phi_{N+1}  = -\omega_N - \frac{K}{3} \left( \sin \phi_{N+1}  -  \sin \phi_{N} \right),
\label{fixed:phase_diff}
\end{align}
where $i = 2,3,\ldots, N$. Then, the constraint in this model reads as
\begin{align}
\sum_{i=1}^{N+1} \phi_i = c_N-c_0 \label{eq:cond1:fix}
\end{align}
The synchronization diagram is shown in Fig.~\ref{kura_n10_fix} for $N=10$ oscillators 
with initial frequencies  $\omega_i =  \{2.5331$, $2.8577$, $2.6978$, $2.0773$, $2.8257$, $1.6062$, $2.7788$, $1.4680$, $1.7622$, $2.0190\}$ and the boundary conditions $c_0 = 0.15$ and $c_N = 2.15$. We notice that the oscillators synchronize with a frequency just below $\omega_0$ at a coupling constant $K_s$. Moreover $K_s$ depends on the initial frequencies and the values of the phases at the fixed ends, i.e. $c_0$ and $c_N$. Increasing the coupling constant, the oscillators remain synchronized in averaged frequency although this value of the average frequency decreases as $K$ increases. Finally at $K=K_c$, the average frequency collapses to zero. The value of $K_c$ depends on $\omega_0$ as well as on the boundary conditions in the phases of the oscillators \cite{me2}.

From the steady states of Eqs.~(\ref{eqn:fixed}) we find that
\begin{align}
K=\frac{a_2}{\sin\phi_2^*-\sin\phi_1^*}, \label{eq:kval:fix}
\end{align}
and
\begin{align}
\sin \phi_1^* - \sin \phi_2^*  + \frac{a_i}{a_2}(\sin \phi_i^*-\sin \phi_1^*) = 0, \; i=3,4\ldots,N, \label{eq:cond2:fix}
\end{align}
where
\begin{align}
a_i=-3\sum_{j=1}^{i-1}\omega_{j}, \; i=2, 3, 4\ldots, N.
\end{align}
The equation to be minimized by using Lagrange multipliers is
\begin{widetext}
\begin{align}
E = \frac{a_2}{\sin \phi_2^*-\sin \phi_1^*} - \sum_{i=3}^{N}
\lambda_i \left[ \sin \phi_1^* - \sin \phi_2^*
 + \frac{a_i}{a_2}(\sin \phi_i^*-\sin \phi_1^*)\right]
+ \lambda_0 \left[\left(\sum_{i=1}^{N} \phi_i^* \right)
-(c_N-c_0)\right] \end{align}
\end{widetext}
Upon minimizing (18), we obtain:
\begin{align}
\sum_{i=1}^{N} \frac{\cos \phi_1^* \cos \phi_2^*}{\cos \phi_i^*} = 0, \label{eq:cond3:fix}
\end{align}
The solution of Eqs.~(\ref{eq:cond1:fix}), (\ref{eq:cond2:fix}) and
(\ref{eq:cond3:fix}) gives $K_c$ provided that we are at stable steady states \begin{table}[!ht]
\caption{Critical $K$ values ($K_c$) for the case fixed ends}
\label{table2}
\begin{tabular}{crr}
\hline
 \multicolumn{1}{c}{N}
    & \multicolumn{2}{c}{$K_c$} \\
 \multicolumn{1}{c}{}
    & \multicolumn{1}{c}{Numerical\footnote{from direct numerical
    simulation of ({\ref{eqn:fixed})}}}
    &  \multicolumn{1}{c}{Analytical\footnote{by solving
    Eqs.~(\ref{eq:cond1:fix}), (\ref{eq:cond2:fix}) and
(\ref{eq:cond3:fix})}} \\
\hline
  3 &  30.343908~~ &  30.343892~~ \\
  6 &  76.993996~~ &  76.993767~~ \\
 10 &  41.127962~~ &  41.127924~~ \\
 12 &  68.656599~~ &  68.656533~~ \\
 15 &  53.059077~~ &  53.059054~~ \\
 25 &  53.968662~~ &  53.967955~~ \\
 50 & 115.546391~~ & 115.546343~~ \\
100 & 80.218507~~ & 80.218133~~ \\
\hline
\end{tabular}
\end{table}
$\phi_i^*$, $i=1,2,3\ldots,N$. In Table~\ref{table2} we show the results obtained for the case of fixed ends, where it is easy to see that the analytic and numeric values of the critical coupling are in a good agreement.

\section{Conclusions}

In this paper we have performed analytic calculations, for a system of coupled oscillators through nearest neighbor coupling with different boundary conditions,  at the transition to complete synchronization. These oscillators are known to synchronize on time average frequencies, forming clusters, increasing the number of oscillators in each,  until they come to complete synchronization. Until this is achieved the oscillators remain time dependent, chaotic or otherwise, as shown by the Lyapunov exponent from Fig.~\ref{lyap_periodic}. We calculated the value of the coupling strength, $K_c$, at complete synchronization. We found that the analytically and numerically calculated values are in a good agreement. In addition we obtain conditions for a complete synchronization to occur.

\acknowledgments

PM thanks the Conselho Nacional de Desenvolvimento Cient\'{\i}fico e Tecnol\'ogico, Brazil, the Third World Academy of Sciences (TWAS-UNESCO Associateship at the Centre of Excellence in the South), and Department of Science and Technology (DST), Government of India. FFF thanks the Conselho Nacional de Desenvolvimento Cient\'{\i}fico e Tecnol\'ogico for financial support and Instituto de F\'{\i}sica Te\'orica for hospitality. HFE thanks the Abdus Salam International Centre for Theoretical Physics for hospitality under the Associateship Scheme and Dr. N. Orlando, the Italian embassy in Riyadh. HAC thanks the Max Planck Institut f\"ur komplexer Systeme for support during part of this work.

\end{document}